\documentclass[12pt]{article}
\usepackage{amsbsy,amsfonts,amssymb}
\newtheorem{sergy-theo}{Theorem}
\newtheorem{sergy-lem}{Lemma}
\newtheorem{sergy-cor}{Corollary}
\newtheorem{sergy-prop}{Proposition}
\date{}
\begin{document}


\title{\protect\vspace*{-1cm}{\bf Time dependence and (non)commutativity\\ 
of symmetries of evolution equations}}

\author{Artur Sergyeyev,\\Mathematical Institute, Silesian University at
Opava,\\
Bezru\v{c}ovo n\'am. 13, 746 01 Opava, Czech Republic\\
E-mail: Artur.Sergyeyev@math.slu.cz,
arthurser@imath.kiev.ua}

\maketitle
\begin{abstract}
We present easily verifiable sufficient conditions of time-independence
and commutativity for local and nonlocal symmetries for a large class of homogeneous
(1+1)-dimensional evolution systems. 
In contrast with the majority of known results, the verification of our conditions
does not require the existence of master symmetry or hereditary recursion 
operator for the system in question.
We also give simple sufficient conditions for the existence of infinite sets of
time-independent symmetries for homogeneous (1+1)-dimensional evolution systems within
slightly modified master  symmetry approach.\looseness=-1
\end{abstract}
\section{Introduction}
Nearly all known today integrable systems are homogeneous with respect to 
some scaling. For such systems no generality is lost in assuming
the homogeneity of symmetries, master symmetries,
recursion operators, etc., 
and this considerably simplifies their finding and study, see e.g.\
\cite{sergy-o}--\cite{sergy-wang}.\looseness=-1 

In the present paper we combine 
this well-known idea with our new results
on the structure of {\em time-dependent} (cf.\ e.g.\
\cite{sergy-sok,sergy-mik1,sergy-s,sergy-mik,sergy-ibrbook} for the
time-in\-de\-pen\-dent case) formal symmetries for a natural generalization 
of the systems, considered in
\cite{sergy-mik1,sergy-s,sergy-muks}, namely, 
for (1+1)-di\-men\-si\-on\-al
nondegenerate weakly diagonalizable (NWD) 
evolution systems with constraints. This enables us to find 
simple sufficient conditions for the commutativity and
time-independence of higher order symmetries and for the existence of infinite number
of such symmetries for {\em homogeneous} NWD systems with constraints. 
Note that the majority of  known \cite{sergy-wang1,sergy-wang,sergy-s} and recently
found, see e.g.\ \cite{sergy-os1,sergy-sokw,sergy-furs}, integrable
evolution systems in (1+1) dimensions fit into this class. Moreover, our results,
unlike the majority of already known ones, are valid for the systems with 
time-dependent coefficients as well, cf.\ e.g.\ \cite{sergy-fu1},  
and are not restricted to scalar equations.
\looseness=-1

Let us stress that 
the proofs and the application of our results involve just
an easy verification of some weight-related conditions and do {\em not} rely on
the existence of a master symmetry or e.g.\ (hereditary) recursion operator. 
Hence, the results of present paper,
except for those on existence of infinitely many symmetries,
can be applied to non-integrable systems as well. 
On the other hand, the simplicity of use makes 
our results particularly helpful in the study of
new integrable systems, for which only a few higher order symmetries and (sometimes) a
`candidate' for the master symmetry are known, but no recursion operator is yet
found. Indeed, we show that the check of a small number of conditions for the 
low order symmetries can replace tedious  
checks, cf.\ \cite{sergy-dor}, that time-in\-de\-pen\-dent symmetries of 
sufficiently high order commute, that a `candidate' for master symmetry
is a nontrivial master symmetry and that its action yields the symmetries
being well-defined (cf.~\cite{sergy-wang} for recursion operators and
{\em local} symmetries) functions of local variables
$x,t,\mathbf{u},\mathbf{u}_{1},\dots$ and of nonlocal variables
$\omega_{\gamma}$, defined below.

Note that, 
unlike \cite{sergy-fu,sergy-oev,sergy-dor}, in order to prove the
existence of infinitely many symmetries we do not make {\em a priori} extra
assumptions, say, about the existence of ``negative" master symmetries
$\boldsymbol\tau_{j}$, $j<0$ \cite{sergy-dor}: 
all we need is a suitable `candidate' $\boldsymbol{\tau}$ for the master symmetry
and a higher order time-independent symmetry.
We also show that in order to verify the commutativity of
{\em all} higher order time-in\-de\-pen\-dent homogeneous 
symmetries at once it suffices to check only a small number of conditions for the
time-in\-de\-pen\-dent symmetries of order lower than two.   
Moreover, 
checks of this kind are almost entirely algorithmic, so computer
algebra software can be readily applied to perform them.\looseness=-2

The paper is organized as follows. In Section \ref{sergy-definitions} we 
give some definitions and facts, being the straightforward extension of those from
\cite{sergy-mik1,sergy-s,sergy-muks} to the case of explicitly time-dependent
evolution systems with constraints. 
In Section 3 we present the sufficient conditions of well-definiteness
of the
symmetries generated by means of master symmetry for the general
evolution systems with constraints. In Section~\ref{sergy-formalsym} 
we define nondegenerate weakly
diagonalizable (NWD) systems with constraints and present some results 
on structure of their formal
symmetries. In Section \ref{sergy-homogen} we find the sufficient conditions 
for commutativity
and time-independence of higher order symmetries and 
for the existence of infinite hierarchies of
time-independent higher order symmetries for homogeneous NWD systems with
constraints.\looseness=-3
\section{Basic definitions and structures}\label{sergy-definitions}
Let us consider a 
system of evolution equations with constraints (cf.\ \cite{sergy-muks})
\begin{equation} \label{sergy-eveq}
\partial \mathbf{u} / \partial t =
\mathbf{F}(x,t,\mathbf{u},\dots,\mathbf{u}_{n'},\ensuremath{\vec \omega})
 \end{equation}
for the vector function $\mathbf{u}=(u^{1},\dots,u^{s})^{T}$. Here
$\mathbf{u}_{j}=\partial^{j}\mathbf{u}/\partial
x^{j}$,$\mathbf{u}_{0}\equiv \mathbf{u}$ and
$\mathbf{F}=(F^{1},\dots, F^{s})^{T}$; $\ensuremath{\vec
\omega}=(\omega_1,\dots,\omega_c)^{T}$; $^{T}$ denotes 
the matrix transposition. The nonlocal variables
$\omega_{\alpha}$ are defined here 
by means of the relations \cite{sergy-muks,sergy-v}\looseness=-1
\begin{eqnarray}
\partial\omega_{\alpha}/\partial
x=X_{\alpha}(x,t,\mathbf{u},\mathbf{u}_{1},\dots,\mathbf{u}_{h},
\ensuremath{\vec\omega}),\label{sergy-omx}\\
\partial\omega_{\alpha}/\partial
t=T_{\alpha}(x,t,\mathbf{u},\mathbf{u}_{1},\dots,\mathbf{u}_{h},
\ensuremath{\vec\omega}).\label{sergy-omt}
\end{eqnarray}
We shall denote by $\Omega$ the set of nonlocal variables
$\omega_{\gamma}$, $\gamma=1,\dots,c$. 

Let $\mathcal{A}_{j,k}(\Omega)$ be the algebra of all
locally analytic scalar functions of
$x,t,\mathbf{u},\mathbf{u}_{1},\allowbreak\dots,\mathbf{u}_{j},
\omega_{1},\dots,\omega_{k}$
with respect to the standard multiplication,
$\mathcal{A}\equiv\mathcal{A}(\Omega)=\bigcup_{k=1}^{c}\bigcup_{j=0}^{\infty}
\mathcal{A}_{j,k}(\Omega)$, and let
$\mathcal{A}_{\mathrm{loc}}=\{f\in\mathcal{A}\mid\partial
f/\partial\ensuremath{\vec \omega}=0\}$ be the subalgebra of {\em local}
functions in $\mathcal{A}$.  Note that we do not exclude the case
$c=\infty$. \looseness=-1 
%
The operators of total $x$- and $t$-derivatives
on $\mathcal{A}$
have the form\looseness=-1
$$
\begin{array}{l}
D\equiv D_x=
{\displaystyle\frac{\partial }{\partial x}} +
\sum\limits_{i=0}^{\infty} \mathbf{u}_{i+1}
{\displaystyle\frac{\partial}{\partial \mathbf{u}_{i}}}
+\sum\limits_{\alpha=1}^{c}X_{\alpha}{\displaystyle
\frac{\partial}{\partial\omega_{\alpha}}},\\
D_{t}=
{\displaystyle\frac{\partial }{\partial t}}+
\sum\limits_{i=0}^{\infty} D^{i}(\mathbf{F})
{\displaystyle\frac{\partial}{\partial \mathbf{u}_{i}}}
+\sum\limits_{\alpha=1}^{c}T_{\alpha}{\displaystyle
\frac{\partial}{\partial\omega_{\alpha}}}.
\end{array}
$$ 
Following \cite{sergy-muks, sergy-v}, we require that $[D_{x},D_{t}]=0$
or, equivalently,
$D_{t}(X_{\alpha})=D_{x}(T_{\alpha})$ for
$\alpha=1,\dots,c$.
\looseness=-1

We shall denote by $\mathop{\rm Im}\nolimits D$ the image of
$\mathcal{A}$ under $D$. Throughout this paper except for Section
\ref{sergy-locality} we make a {\em blanket assumption} that the kernel of $D$ in
$\mathcal{A}$ consists solely of functions of $t$.
\looseness=-2

Consider the set
$\mathrm{Mat}_{p}(\mathcal{A})[\![D^{-1}]\!]$ of {\em formal series} in
powers of $D$ of the form
$\mathfrak{H}
=\sum
_{j=-\infty}^{q} h_{j}D^{j}$, 
where $h_{j}$ are $p\times p$ matrices with 
entries from $\mathcal{A}$, cf.\ e.g.\ \cite{sergy-mik1,sergy-s}. We shall
write for short $\mathcal{A}[\![D^{-1}]\!]$ instead of
$\mathrm{Mat}_{1}(\mathcal{A})[\![D^{-1}]\!]$.
\looseness=-1

The greatest $m\in\mathbb{Z}$ such that $h_{m}\neq 0$ is called
the {\em degree} of 
$\mathfrak{H}
\in\mathrm{Mat}_{p}(\mathcal{A})[\![D^{-1}]\!]$ and is denoted as $m
=\deg\mathfrak{H}$. We assume 
that $\deg 0=-\infty$, cf.\ e.g.\ \cite{sergy-o}. The formal series 
$\mathfrak{H}$ of
degree $m$ is called {\em non\-de\-ge\-ne\-rate} \cite{sergy-s}, 
if $\det h_{m}\neq
0$. For $\mathfrak{H}=\sum
_{j=-\infty}^{m} h_{j}D^{j}\in\mathcal{A}[\![D^{-1}]\!]$, $h_m\neq 0$, 
its {\em residue} and {\em logarithmic residue} are defined as
$\mathop{\rm
res}\nolimits\mathfrak{H}=h_{-1}$ and $\mathop{\rm
res}\nolimits\ln\mathfrak{H}=h_{m-1}/h_{m}$
\cite{sergy-mik1,sergy-s}.\looseness=-1

The set $\mathrm{Mat}_{p}(\mathcal{A})[\![D^{-1}]\!]$
is an algebra under the multiplication law, given by the
``generalized Leibniz rule", cf.\ \cite{sergy-o}, 
\looseness=-1
$$ a D^{i}\circ b D^{j} =a
\sum\limits_{q=0}^{\infty}{\displaystyle \frac{i(i-1)\cdots
(i-q+1)}{q!}}D^{q}(b)D^{i+j-q} $$
for monomials $a D^{i}$, $b D^{j}$, $a,b\in\mathrm{Mat}_{p}(\mathcal{A})$,
and extended by linearity to the whole
$\mathrm{Mat}_{p}(\mathcal{A})[\![D^{-1}]\!]$. 
The commutator $\left[ \mathfrak{A}, \mathfrak{B}
\right]=\mathfrak{A} \circ \mathfrak{B}- \mathfrak{B} \circ
\mathfrak{A}$ makes $\mathrm{Mat}_{p}(\mathcal{A})[\![D^{-1}]\!]$ 
into a Lie algebra.
Below we omit
$\circ$ if this is not confusing.\looseness=-1

$\mathbf{G}\in\mathcal{A}^{s}$ is called, 
see e.g.\ \cite{sergy-o,sergy-fokas,sergy-blaszak}, a {\em symmetry} for
(\ref{sergy-eveq})--(\ref{sergy-omt}), if
\begin{equation}\label{sergy-sym}
\partial\mathbf{G}/\partial t+[\mathbf{F},\mathbf{G}]=0,
\end{equation}
where $[\cdot,\cdot]$ is the 
Lie bracket $[\mathbf{K},\mathbf{H}]=
\mathbf{H}'[\mathbf{K}]-\mathbf{K}'[\mathbf{H}]$. The
directional 
derivative of any (smooth) function
$f\in\mathcal{A}^{q}$ along 
$\mathbf{H}\in\mathcal{A}^{s}$ is defined here as
$f'[\mathbf{H}]=(d
f(x,t,\mathbf{u}+\epsilon\mathbf{H},\mathbf{u}_{1}+\epsilon
D_{x}(\mathbf{H}),\dots)/ d\epsilon)_{\epsilon=0}$. 
Extending the technique of \cite{sergy-muks} to the case of
time-dependent systems (\ref{sergy-eveq})--(\ref{sergy-omt}), we
can easily show that for any $f\in\mathcal{A}$ 
we have $f'\in\mathcal{A}[\![D^{-1}]\!]$.

For any $f\in\mathcal{A}^{q}$ we shall define its {\em formal
order} as $\mathop{\rm ford}\nolimits f=\deg f'$.
This naturally generalizes the notion of order for local functions,
cf.\ e.g.\ \cite{sergy-o,sergy-s}.\looseness=-1

Let ${S}_{F}(\mathcal{A})$ be the set of all symmetries
$\mathbf{G}\in\mathcal{A}^{s}$ for (\ref{sergy-eveq})--(\ref{sergy-omt}),
${S}_{F}^{(k)}(\mathcal{A})=\{\mathbf{G}\in{S}_{F}(\mathcal{A})\mid\mathop{\rm
ford}\nolimits\mathbf{G}\leq k\}$, $\mathrm{Ann}_{F}(\mathcal{A})=\{\mathbf{G}\in
{S}_{F}(\mathcal{A})\mid\partial\mathbf{G}/\partial t=0\}$. 
In general, for $\mathcal{A}\neq\mathcal{A}_{\mathrm{loc}}$ neither
$\mathcal{A}^{s}$ nor ${S}_{F}(\mathcal{A})$ are closed under the Lie bracket,
but if $[\mathbf{P},\mathbf{Q}]\in\mathcal{A}^{s}$ for some
$\mathbf{P},\mathbf{Q}\in S_{F}(\mathcal{A})$, then we have
$[\mathbf{P},\mathbf{Q}]\in S_{F}(\mathcal{A})$.

A formal series $\mathfrak{R}
=\sum
_{j=-\infty}^{r}\eta_{j}D^{j}
\in\mathrm{Mat}_{s}(\mathcal{A})[\![D^{-1}]\!]$
is called \cite{sergy-o,sergy-mik1,sergy-muks} the {\em formal symmetry} of rank
$m$ for (\ref{sergy-eveq}) (or, rather, for
(\ref{sergy-eveq})--(\ref{sergy-omt})), provided
\begin{equation}\label{sergy-wfs-def} \deg
(D_{t}(\mathfrak{R})-[\mathbf{F}', \mathfrak{R}])\leq \deg
\mathbf{F}'+\deg\mathfrak{R}-m.
\end{equation}
The derivative $D_{t}(\mathfrak{R})$ is defined here as 
$D_{t}(\mathfrak{R})=\sum
_{j=-\infty}^{r}D_{t}(\eta_{j})D^{j}$.

The set $FS_{F}^{(q)}(\mathcal{A})$ of all formal symmetries of rank not
lower than $q$ of system (\ref{sergy-eveq})--(\ref{sergy-omt}) is a Lie
algebra, because 
for the formal symmetries $\mathfrak{P}$ and $\mathfrak{Q}$ of ranks 
$p$ and $q$ 
we have $[\mathfrak{P},\mathfrak{Q}]\in FS_{F}^{(r)}(\mathcal{A})$ for
$r=\min(p,q)$, cf.\  \cite{sergy-s}.\looseness=-1

Eq.(\ref{sergy-sym}) is well known to be nothing but the compatibility 
condition for
(\ref{sergy-eveq}) and $\partial\mathbf{u}/\partial\tau=\mathbf{G}$. 
Provided $\mathbf{G}\in\mathcal{A}^{s}$, we have
$\partial(\partial \mathbf{u}/\partial\tau)\partial
t=D_{t}(\mathbf{G})$ and $\partial(\partial \mathbf{u}/\partial
t)\partial\tau=\mathbf{F}'[\mathbf{G}]$.
Hence Eq.(\ref{sergy-sym}) may be
rewritten as $D_{t}(\mathbf{G})=\mathbf{F}'[\mathbf{G}]$. 

Let $\mathbf{F}'\equiv\sum\limits_{i=-\infty}^{n}\phi_{i}D^i$ and
$n_0=\left\{\begin{array}{l} 1-j,\,\mbox{if}\,\phi_i=\phi_i(x,t), 
i=n-j,\dots,n,\\ 2\,\,\mbox{otherwise.} \end{array} \right.$ 

Since $D_{t}(\mathbf{G})=\mathbf{F}'[\mathbf{G}]$ implies 
$D_{t}(\mathbf{G}')-[\mathbf{F}',\mathbf{G}']-
\mathbf{F}''[\mathbf{G}]=0$, and $\deg \mathbf{F}''[\mathbf{G}]\leq \deg
\mathbf{F}'+n_0-2$, we readily see that 
$\mathbf{G}'\in {FS}_{F}^{(\mathop{\rm
ford}\nolimits\mathbf{G}-n_0+2)}(\mathcal{A})$.

\section{Action of master symmetries on time-independent
symmetries}\label{sergy-locality} As we have already mentioned above, for
$\mathbf{P},\mathbf{Q}\in\mathcal{A}^{s}$ in general
$[\mathbf{P},\mathbf{Q}]\not\in\mathcal{A}^{s}$. In particular, when we 
repeatedly commute a master symmetry $\boldsymbol\tau\in
\mathcal{A}^{s}$ with some time-independent symmetry
$\mathbf{Q}\in\mathrm{Ann}_{F}(\mathcal{A})$, it is by no means obvious that
$\mathbf{Q}_{i}=\mathop{\rm ad}\nolimits
_{\boldsymbol{\tau}}^{i}(\mathbf{Q})=[\boldsymbol\tau,\mathbf{Q}_{i-1}]$ belong to
$\mathcal{A}^{s}$, except for the case $\mathcal{A}=\mathcal{A}_{\mathrm{loc}}$. 
In some cases we can make the conditions
$[\boldsymbol{\tau},\mathbf{Q}_{i}]\in\mathcal{A}^{s}$ or
$[\mathbf{P},\mathbf{Q}]\in\mathcal{A}^{s}$ hold by 
introducing new nonlocal variables
$\tilde{\omega}_{\kappa}$ and thus replacing $\mathcal{A}$ by a larger algebra
$\tilde{\mathcal{A}}$.
But in order that $[\mathbf{P},\mathbf{Q}]\in\mathcal{A}^{s}$ for
$\mathbf{P},\mathbf{Q}\in\mathcal{A}^{s}$ it obviously suffices 
to require 
that $\omega_{\mu}'[\mathbf{P}]\in\mathcal{A}$ for those
$\omega_{\mu}$ on which $\mathbf{Q}$ actually depends and
$\omega_{\nu}'[\mathbf{Q}]\in\mathcal{A}$ for those
$\omega_{\nu}$ on which $\mathbf{P}$ actually
depends, cf.\ Ch.~6 in \cite{sergy-v}.\looseness=-1

 Moreover, we have
\begin{sergy-prop}\label{sergy-prop4}
Let
$\boldsymbol\tau,\mathbf{Q}\in
\mathcal{A}^{s}$,
$\omega'_{\gamma}[\mathbf{Q}]\in\mathcal{A}$ and
$\omega'_{\gamma}[\boldsymbol{\tau}]\in\mathcal{A}$ for
$\gamma=1,\dots,c$.
Then
$\mathbf{Q}_l=\mathop{\rm
ad}\nolimits_{\boldsymbol{\tau}}^{l}(\mathbf{Q})\in\mathcal{A}^{s}$ for all
$l=1,2,\dots$.
\end{sergy-prop}
{\em Proof.} Let us use induction. To prove that
$[\boldsymbol{\tau},\mathbf{Q}_{l}]\in\mathcal{A}^{s}$, if
$\mathbf{Q}_{l}=[\boldsymbol{\tau},\mathbf{Q}_{l-1}]
\in\mathcal{A}^{s}$, it 
suffices to prove that
$\omega'_{\nu}([\boldsymbol{\tau},\mathbf{Q}_{l-1}])\in\mathcal{A}$
for all $\omega_{\nu}$ which $\boldsymbol{\tau}$ 
depends on and that
$\omega_{\delta}'[\boldsymbol{\tau}]\in\mathcal{A}$
for all $\omega_{\delta}$ which
$[\boldsymbol{\tau},\mathbf{Q}_{l-1}]$ 
depends on.
As $\omega'_{\nu}([\boldsymbol{\tau},\mathbf{Q}_{l-1}])=
(\omega'_{\nu}[\mathbf{Q}_{l-1}])'[\boldsymbol{\tau}]-
(\omega'_{\nu}[\boldsymbol{\tau}])'[\mathbf{Q}_{l-1}]$,
it suffices that
$\omega'_{\gamma}[\boldsymbol{\tau}]\in\mathcal{A}$ for all
$\omega_{\gamma}$, which $[\boldsymbol{\tau},\mathbf{Q}_{l-1}]$
and $\omega'_{\nu}[\mathbf{Q}_{l-1}]$ depend on, and
$\omega'_{\kappa}[\mathbf{Q}_{l-1}]\in\mathcal{A}$ for all
$\omega_{\kappa}$ which $\boldsymbol{\tau}$ and
$\omega'_{\nu}[\boldsymbol{\tau}]$
depend on, in order that $[\boldsymbol{\tau},\mathbf{Q}_{l}]\in\mathcal{A}^{s}$.
$\square$\looseness=-2

It appears that nearly all known master symmetries of integrable systems 
(\ref{sergy-eveq})--(\ref{sergy-omt}) satisfy the conditions of
Proposition~\ref{sergy-prop4} for a suitably chosen set $\Omega$  of nonlocal 
variables $\omega_{\gamma}$, so their action indeed yields the
symmetries from
$\mathcal{A}^{s}$. For instance, if
$\partial\mathbf{F}/\partial\vec{\omega}=0$ and
$\mathcal{A}=\mathcal{A}(\Omega_{\mathrm{UAC},F})$, then
by virtue of the results of \cite{sergy-v} 
Proposition~\ref{sergy-prop4} holds true for any
$\boldsymbol{\tau},\mathbf{Q}\in S_{F}(\mathcal{A})$. 
%
Here $\Omega_{\mathrm{UAC},F}$ is the set of all nonlocal variables
$\omega_{\gamma}$ associated with the universal abelian covering
(see \cite{sergy-v} for its definition) over~(\ref{sergy-eveq}).
Let us stress that Proposition~\ref{sergy-prop4} is valid for any 
$\boldsymbol\tau$ and $\mathbf{Q}$ meeting the relevant requirements, 
no matter whether $\boldsymbol\tau$ is a master symmetry and $\mathbf{Q}$
is a symmetry for (\ref{sergy-eveq})--(\ref{sergy-omt}).
\looseness=-2

Note that Proposition~\ref{sergy-prop4} is obviously valid for more
general systems of PDEs with constraints than
(\ref{sergy-eveq})--(\ref{sergy-omt}), if we suitably redefine for them
the Lie bracket, the directional derivative and the algebra
$\mathcal{A}$.\looseness=-1

\section{The structure of formal symmetries for NWD systems}\label{sergy-formalsym}
Consider
a particular class of evolution systems with constraints
(\ref{sergy-eveq})--(\ref{sergy-omt}) such that
$n\equiv\mathop{\rm ford}\nolimits\mathbf{F}\geq 2$ and the
leading coefficient $\Phi$ of the formal series $\mathbf{F}'$
(i.e., $\mathbf{F}'\equiv\Phi D^{n}+\dots$) has $s$ distinct eigenvalues 
$\lambda_i$ and can
be diagonalized by means of a matrix
$\Gamma=\Gamma(x,t,\mathbf{u},\dots,\mathbf{u}_{n'},\ensuremath{\vec
\omega})$, i.e., the matrix $\Lambda=\Gamma \Phi \Gamma^{-1}$ is
diagonal, cf.\ \cite{sergy-mik1,sergy-s}.  
For these systems 
there exists a unique formal series
$\mathfrak{T}=\Gamma+
\Gamma\sum_{j=1}^{\infty}\Gamma_{j}
D^{-j}\in
\mathrm{Mat}_{s}(\mathcal{A})[\![D^{-1}]\!]$ such that all coefficients 
of the formal series
$\mathfrak{V}=\mathfrak{T}\mathbf{F}'\mathfrak{T}^{-1} +
(D_{t}(\mathfrak{T})) {\mathfrak{T}}^{-1}$
are diagonal matrices and the diagonal entries of matrices
$\Gamma_j$,$j=1,2,\dots$, are equal to zero.
The proof is essentially the same as for
Proposition~3.1 from \cite{sergy-mik1}. 
We shall call the systems
with constraints (\ref{sergy-eveq})--(\ref{sergy-omt}) having the above properties
and such that $\det\Phi\neq 0$ {\it nondegenerate weakly diagonalizable (NWD)}.
\looseness=-2
%
Note that when $\mathbf{u}$ is scalar, i.e., $s=1$, any system
(\ref{sergy-eveq})--(\ref{sergy-omt}) 
with $n\equiv\mathop{\rm ford}\nolimits\mathbf{F}\geq 2$ obviously is an NWD system
with constraints, having $\mathfrak{T}=1$ and
$\mathfrak{V}=\mathbf{F}'$.\looseness=-1

Below in this section (\ref{sergy-eveq})--(\ref{sergy-omt}) will be an NWD
system with constraints.
\looseness=-1

Eq.(\ref{sergy-wfs-def}) yields \cite{sergy-mik1,sergy-s} 
$\deg(D_{t}(\tilde{\mathfrak{R}})-
[\mathfrak{V},\tilde{\mathfrak{R}}])\leq\deg\mathfrak{V}
+\deg\tilde{\mathfrak{R}}-m$, where
$\tilde{\mathfrak{R}}=
\mathfrak{T}\mathfrak{R}\mathfrak{T}^{-1}$, 
whence
we find (cf.\ \cite{sergy-sok,sergy-s,sergy-mik}) that  any
$\mathfrak{R}\in{FS}^{(n+1)}_{F}(\mathcal{A})$ 
can be represented in the form\looseness=-1
\begin{equation}\label{sergy-leadfs2} \begin{array}{l}
\mathfrak{R}=\mathfrak{T}^{-1}\left(\sum\limits_{j=r-n+1}^{r}
c_{j}(t) \mathfrak{V}^{j/n}\right)\mathfrak{T}+{\displaystyle\frac{1}{n}}
\mathfrak{T}^{-1}\left(D^{-1}\left(\dot c_{r}(t)
\Lambda^{-1/n}\right.\right. \\[5mm] \left.\left. 
\quad - r c_{r}(t)D_{t}(\Lambda^{-1/n})\right)
\right)\mathfrak{V}^{\frac{r-n+1}{n}}\mathfrak{T}
+\mathfrak{N},\, \deg\mathfrak{N}<r-n+1.
\end{array}
\end{equation}
Likewise, for $\mathfrak{R}\in{FS}^{(m)}_{F}(\mathcal{A})$ with $m=2,\dots,
n$ we have
\begin{equation}\label{sergy-leadfs2a} \begin{array}{l}
\mathfrak{R}=\mathfrak{T}^{-1}\left(\sum\limits_{j=r-m+2}^{r}
c_{j}(t) \mathfrak{V}^{j/n}\right)\mathfrak{T}
+\mathfrak{N},\, \deg\mathfrak{N}<r-m+2.
\end{array}
\end{equation}
Here $r=\deg\mathfrak{R}$, $\mathfrak{N}=
b D^{\nu}+\cdots\in\mathrm{Mat}_{s}(\mathcal{A})[\![D^{-1}]\!]$  is some formal
series, $\nu=r-n$ in (\ref{sergy-leadfs2}) and $\nu=r-m+1$ in (\ref{sergy-leadfs2a});
$c_{j}(t)$ and $\Gamma b \Gamma^{-1}$ are diagonal $s
\times s$ matrices;  
for $\mathfrak{V}\equiv\mathop{\rm
diag}\nolimits(\mathfrak{V}_{1},\dots,\mathfrak{V}_{s})$,
$\mathfrak{V}_{i}\in\mathcal{A}[\![D^{-1}]\!]$,
we set $\mathfrak{V}^{j/n}=\mathop{\rm
diag}\nolimits(\mathfrak{V}_{1}^{j/n},\dots,\mathfrak{V}_{s}^{j/n})$
\cite{sergy-mik1};  dot stands for the {\em partial} derivative w.r.t.~$t$.

In this section we assume that any function $\tilde{h}+a(t)$, where $a(t)$ is an
arbitrary function of $t$, can be taken for 
$D^{-1}(h)$, if $h=D(\tilde{h})$ and $h, \tilde{h}\in\mathcal{A}$.
\looseness=-1

For $m=2,\dots, n+1$ Eqs.~(\ref{sergy-leadfs2}),
(\ref{sergy-leadfs2a}) represent a general solution of
(\ref{sergy-wfs-def}) for any NWD system with
constraints (\ref{sergy-eveq})--(\ref{sergy-omt}).  
Hence, if at least one entry
of the matrix 
$(\dot c_{r}(t)\Lambda^{-1/n}-r c_{r}(t)D_{t}(\Lambda^{-1/n}))$
does not belong to $\mathop{\rm Im}\nolimits D$, 
then 
(\ref{sergy-eveq})--(\ref{sergy-omt}) has no formal
symmetries from $FS_{F}^{(n+1)}(\mathcal{A})$ with a
given $c_{r}(t)$.\looseness=-2 

For any ${\mathfrak{P}}\equiv\mathfrak{T}^{-1}c_{p} (t)
\mathfrak{V}^{p/n}\mathfrak{T}+\cdots$ and
${\mathfrak{Q}}\equiv\mathfrak{T}^{-1}d_{q} (t)
\mathfrak{V}^{q/n}\mathfrak{T}+\cdots$ we have 
\looseness=-2
\begin{equation}\label{sergy-leadcomm}
\lbrack{\mathfrak{P}},{\mathfrak{Q}}\rbrack=\mathfrak{T}^{-1}
(1/n)(p c_{p}(t)\dot
d_{q}(t)-q d_{q}(t)\dot c_{p}(t))\mathfrak{V}^{\frac{p+q-n}{n}}
\mathfrak{T}+\mathfrak{K}
\end{equation}
by virtue of (\ref{sergy-leadfs2}), provided $\mathfrak{P},\mathfrak{Q}\in
FS_{F}^{(n+1)}(\mathcal{A})$. Here
$\mathfrak{K}\in\mathrm{Mat}_{s}(\mathcal{A})[\![D^{-1}]\!]$ is some 
formal series, $\deg\mathfrak{K}<p+q-n$.

Let $\mathbf{P},
\mathbf{Q}\in\mathcal{A}^{s}$,
$\mathbf{R}\equiv\lbrack{\mathbf{P}},{\mathbf{Q}}\rbrack$. 
Then $\mathbf{R}'
=\mathbf{Q}''[\mathbf{P}]-\mathbf{P}''[\mathbf{Q}]-
[\mathbf{P}',\mathbf{Q}']$. If $\mathbf{P},\mathbf{Q}\in
S_{F}(\mathcal{A})$, then (\ref{sergy-wfs-def}) and (\ref{sergy-leadfs2}) for
$\mathfrak{R}=\mathbf{P}'$ and $\mathfrak{R}=\mathbf{Q}'$ imply 
$\deg\mathbf{P}''[\mathbf{Q}]\leq p+n_0-2<p+q-n$ and 
$\deg\mathbf{Q}''[\mathbf{P}]\leq q+n_0-2<p+q-n$ for $p,q>n+n_0-2$,
$p\equiv\mathop{\rm ford}\nolimits\mathbf{P}$, $q\equiv\mathop{\rm
ford}\nolimits\mathbf{Q}$. This result and 
(\ref{sergy-leadcomm}) for $\mathfrak{P}=\mathbf{P}'$, $\mathfrak{Q}=\mathbf{Q}'$
yield\looseness=-1
\begin{equation}\label{sergy-leadcomm1}
\lbrack{\mathbf{P}},{\mathbf{Q}}\rbrack'=-\mathfrak{T}^{-1}
(1/n)(p c_{p}(t)\dot
d_{q}(t)-q d_{q}(t)\dot c_{p}(t))\mathfrak{V}^{\frac{p+q-n}{n}}
\mathfrak{T}+\tilde{\mathfrak{K}},
\end{equation}
where $\tilde{\mathfrak{K}}\in\mathrm{Mat}_{s}(\mathcal{A})[\![D^{-1}]\!]$ 
is some formal series, $\deg\tilde{\mathfrak{K}}<p+q-n$. \looseness=-1

So, if $\mathbf{P},\mathbf{Q}\in S_{F}(\mathcal{A})$, $p,q>n+n_0-2$,
then $\mathop{\rm ford}\nolimits\mathbf{R}\leq p+q-n$. If
$\mathbf{R}\in\mathcal{A}^{s}$, then
$\mathbf{R}\in S_{F}^{(p+q-n)}(\mathcal{A})$, and $\mathbf{R}\in
S_{F}^{(p+q-n-1)}(\mathcal{A})$, if $p c_{p}(t)\dot d_{q}(t)=q d_{q}(t)\dot
c_{p}(t)$.
\looseness=-1 

Let 
(\ref{sergy-eveq})--(\ref{sergy-omt}) have a nondegenerate
formal symmetry $\mathfrak{R}\in\mathrm{Mat}_{s}(\mathcal{A})[\![D^{-1}]\!]$, 
$r\equiv\deg\mathfrak{R}\neq 0$, 
of rank $q>n$. Then $D_{t}(\rho_{j}^{a})\in\mathop{\rm
Im}\nolimits D$, i.e., $\rho_{j}^{a}$   
are conserved densities, for $a=1,\dots,s$ and $j=-1,0,\dots,q-n-2$,
where $\rho_{0}^{a}=\mathop{\rm
res}\nolimits\ln((\mathfrak{T}\mathfrak{R}
\mathfrak{T}^{-1})^{1/r})_{aa}$ and
$\rho_{j}^{a}=\mathop{\rm res}\nolimits((\mathfrak{T}
\mathfrak{R}\mathfrak{T}^{-1})^{j/r})_{aa}$ for $j\neq 0$, cf.\ \cite{sergy-mik1}.
For $n_0<2$ we have $\rho_{j}^{a}\in\mathop{\rm Im}\nolimits D$ for all
$a=1,\dots,s$ and $j=-1,0,\dots,-n_0$.\looseness=-1
\begin{sergy-prop}\label{sergy-prop3}
Let an NWD system with constraints 
(\ref{sergy-eveq})--(\ref{sergy-omt}) have a nondegenerate formal
symmetry $\mathfrak{R} 
\in\mathrm{Mat}_{s}(\mathcal{A})[\![D^{-1}]\!]$, 
$r\equiv\mathop{\rm deg}\nolimits\mathfrak{R}\neq 0$, 
$q\equiv\mathop{\rm rank}\nolimits\mathfrak{R}>n$; 
let for 
$a=1,\dots,s$  there exist $m_{a}\in
\{-1,1,2,\dots$, $\min(n-2,q-n-2)\}$ such that $m_{a}\neq 0$,
$\rho_{m_{a}}^{a}\not\in\mathop{\rm Im}\nolimits D$ and
$\rho_{j}^{a}\in\mathop{\rm Im}\nolimits D$ for 
$j=-1,1\dots,m_{a}-1$, $j\neq 0$. 
Then for each $\mathfrak{P}\in
FS_{F}^{(m+n+2)}(\mathcal{A})$, $m=\max\limits_{a} m_a$, there exists
a constant $s\times s$ diagonal matrix $c$ such that
$\mathfrak{P}=\mathfrak{T}^{-1}c
\mathfrak{R}^{p/r}\mathfrak{T}+\cdots$, $p\equiv\deg\mathfrak{P}$.\looseness=-2
\end{sergy-prop}
{\em Proof.} 
Since $\mathfrak{R}\in FS_{F}^{(n+1)}(\mathcal{A})$, by (\ref{sergy-leadfs2}) we have
$\mathfrak{R}=\mathfrak{T}^{-1} h(t)\mathfrak{V}^{r/n}\mathfrak{T}+\cdots$.
For any $\mathfrak{P}\in FS_{F}^{(n+1)}(\mathcal{A})$ 
we can (cf.\ \cite{sergy-sok,sergy-ibrbook} and
(\ref{sergy-leadfs2})) represent
$\tilde{\mathfrak{P}}\equiv\mathfrak{T}\mathfrak{P}\mathfrak{T}^{-1}$ as
\looseness=-1
\begin{equation}\label{sergy-leadfs3} \begin{array}{l}
\tilde{\mathfrak{P}}
=\hspace*{-3mm}\sum\limits_{j=p-n+1}^{p}\hspace{-2mm} c_{j}(t)
\tilde{\mathfrak{R}}^{j/r}+{\displaystyle\frac{1}{n}}\left(D^{-1}\left(\dot
c_{p}(t)(h(t))^{n/r}\rho_{-1}\right)\right)\tilde{\mathfrak{R}}^{\frac{p-n+1}{r}}
+\tilde{\mathfrak{N}}. \end{array} \end{equation} 
Here
$\tilde{\mathfrak{N}}\equiv\sum\limits_{j=-\infty}^{p-n}\tilde{b}_{j}
D^{j}\in\mathrm{Mat}_{s}(\mathcal{A})[\![D^{-1}]\!]$, 
$c_{j}(t)$, $h(t)$, $\tilde{b}_{p-n}$ are diagonal $s \times s$
matrices, 
$\rho_{-1}\equiv\mathop{\rm
diag}\nolimits(\rho_{-1}^{1},\dots,\allowbreak\rho_{-1}^{s})$,
$\tilde{\mathfrak{R}}=\mathfrak{T}\mathfrak{R}\mathfrak{T}^{-1}$; 
the fractional powers $\tilde{\mathfrak{R}}^{j/r}$ are defined 
so that their first $r$ coefficients are diagonal, cf.\
\cite{sergy-mik1,sergy-s}.\looseness=-2
 
For $\mathfrak{P}\in FS_{F}^{(d)}(\mathcal{A})$ we have 
$\deg(D_{t}(\tilde{\mathfrak{P}})-[\mathfrak{V},\tilde{\mathfrak{P}}])\leq n+p-d$,
and thus 
$\deg(D_{t}(\tilde{\mathfrak{P}}_{i})
-[\mathfrak{V},\tilde{\mathfrak{P}}_{i}])\leq n+p+i-\min(q,d)$
for $\tilde{\mathfrak{P}}_{i}\equiv\tilde{\mathfrak{P}}\tilde{\mathfrak{R}}^{i/r}$.
Hence, for $-p-2<i<\min(q,d)-n-p-1$ we have $\mathop{\rm
res}\nolimits(D_{t}(\tilde{\mathfrak{P}}_{i})-
[\mathfrak{V},\tilde{\mathfrak{P}}_{i}])=0$. 

Let us plug (\ref{sergy-leadfs3}) into this equality for
$-p-2<i<\min(q,d,2n)-n-p-1$ and break it into
$s$ scalar equations. Since 
$\mathop{\rm res}\nolimits([\mathfrak{V},\tilde{\mathfrak{P}}_{i}])_{aa}
\in\mathop{\rm Im}\nolimits D$ by Adler's formula, see e.g.\ \cite{sergy-s}, and
$D_{t}(\rho_{j+i}^a)\in\mathop{\rm Im}\nolimits D$ by assumption, we easily  find
that for any $\mathfrak{P}\in FS_{F}^{(m+n+2)}(\mathcal{A})$ we have 
$(\dot c_{p}(t))_{aa}\rho^a_{m_{a}}=0$ modulo the terms from 
$\mathop{\rm Im}\nolimits D$ for all $a=1,\dots,s$. So, $\dot c_{p}(t)=0$, and the
result follows. $\square$\looseness=-1
\begin{sergy-cor}\label{sergy-cor2}
Under the assumptions of Proposition \ref{sergy-prop3},
for any $\mathbf{G}\in
S_{F}(\mathcal{A})$, $k\equiv\mathop{\rm ford}\nolimits\mathbf{G}\geq m+n+n_0$,
we have
$\mathbf{G}'=\mathfrak{T}^{-1}c
\mathfrak{R}^{k/r}\mathfrak{T}+\cdots$, 
where $c$ is a constant $s\times s$ diagonal matrix. 
\end{sergy-cor}
\section{Symmetries of homogeneous NWD systems}\label{sergy-homogen}
Let (\ref{sergy-eveq})--(\ref{sergy-omt}) possess a scaling symmetry
$\mathbf{D}=\alpha t\mathbf{F}+x\mathbf{u}_1+\beta\mathbf{u}$,
where $\beta=\mathrm{diag}(\beta_1,\dots,\beta_s)$ is a diagonal matrix,
$\alpha,\beta_j={\rm const}$, and let the determining equations (\ref{sergy-omx}),
(\ref{sergy-omt}) for $\omega_{\gamma}$, $\gamma=1,\dots,c$, be homogeneous 
with respect to~$\mathbf{D}$. Then we shall call the evolution
system with constraints (\ref{sergy-eveq})--(\ref{sergy-omt}) {\em homogeneous}
w.r.t.~$\mathbf{D}$, cf.\ e.g.\ \cite{sergy-os1,sergy-wang1,sergy-wang,
sergy-v}. 
If a formal vector field $\mathbf{G}\partial/\partial\mathbf{u}$ 
is homogeneous of weight $\kappa$ w.r.t.~$\mathbf{D}$, then we shall say for short  
that $\mathbf{G}\in\mathcal{A}^{s}$ itself is homogeneous of weight
$\kappa$ and write
$\mathop{\rm wt}\nolimits(\mathbf{G})=\kappa$. 

For homogeneous systems
(\ref{sergy-eveq})--(\ref{sergy-omt}) there usually exists a basis 
in $S_{F}(\mathcal{A})$ made of homogeneous
symmetries, and hence the requirement of homogeneity of $\mathbf{P},\mathbf{Q}$
and
$\boldsymbol\tau$ below is by no means restrictive. So, the phrase like ``for all
(homogeneous)
$\mathbf{H}\in\mathcal{M}$ the condition $P$ is true" below
means that there exists a basis in $\mathcal{M}$ such that all its
elements are homogeneous w.r.t.~$\mathbf{D}$, and for all 
of them the
condition
$P$ holds true. We have an obvious\looseness=-1 
\begin{sergy-lem}\label{sergy-lem1}
Let (\ref{sergy-eveq})--(\ref{sergy-omt}) be a homogeneous
system with constraints, and homogeneous $\mathbf{P},\mathbf{Q}\in
S_{F}(\mathcal{A})$ be such that $[\mathbf{P},\mathbf{Q}]\in\mathcal{M}$, where
$\mathcal{M}$ is a subspace of $\mathcal{A}^{s}$. 
Suppose that $\mathop{\rm
wt}\nolimits(\mathbf{G})\neq\mathop{\rm
wt}\nolimits([\mathbf{P},\mathbf{Q}])=\mathop{\rm
wt}\nolimits(\mathbf{P})+\mathop{\rm wt}\nolimits(\mathbf{Q})$ for all (homogeneous)
$\mathbf{G}\in S_{F}^{(p+q)}(\mathcal{A})\cap\mathcal{M}$,
$p\equiv\mathop{\rm ford}\nolimits\mathbf{P}$,
$q\equiv\mathop{\rm ford}\nolimits\mathbf{Q}$. 
Then $[\mathbf{P},\mathbf{Q}]=0$.
\looseness=-1
\end{sergy-lem}

This result, as well as other results below, allows to prove the
commutativity for large {\em families} of symmetries at once.
Examples below show that we can usually choose the subspaces like $\mathcal{M}$
large enough so that the condition $[\mathbf{P},\mathbf{Q}]\in\mathcal{M}$ 
can be verified for all symmetries in the family 
without actually computing
$[\mathbf{P},\mathbf{Q}]$. On the other hand, by proper choice of
these subspaces we can considerably reduce the number of weight-re\-lated conditions
to be verified, and thus make the application of our results truly
efficient. 
\looseness=-1

Below in this section 
we assume that 
(\ref{sergy-eveq})--(\ref{sergy-omt}) is a homogeneous NWD system with
constraints and 
$\mathbf{P},\mathbf{Q}\in
S_{F}(\mathcal{A})$ are its {\em homogeneous} symmetries,
$p\equiv\mathop{\rm ford}\nolimits\mathbf{P}$,
$q\equiv\mathop{\rm ford}\nolimits\mathbf{Q}$. 
Note that if $p,q>n+n_0-2$,
then by (\ref{sergy-leadcomm1}) we should verify the conditions of
Lemma~\ref{sergy-lem1} only for $\mathbf{G}\in
S_{F}^{(p+q-n)}(\mathcal{A})\cap\mathcal{M}$ (for
$\mathbf{G}\in S_{F}^{(p+q-n-1)}(\mathcal{A})\cap\mathcal{M}$, if in addition
$p c_{p}(t)\dot d_{q}(t)-q d_{q}(t)\dot c_{p}(t)=0$).\looseness=-2
\subsection{Commutativity and time dependence of
symmetries}\label{sergy-homogen-com}
\begin{sergy-cor}\label{sergy-corann}
Let $\alpha\neq 0$, $\partial\Phi/\partial t=0$ and $\partial
X_{\gamma}/\partial t=\partial T_{\gamma}/\partial t=0$, $\gamma=1,\dots,c$.
Let homogeneous $\mathbf{P},\mathbf{Q}\in
\mathrm{Ann}_{F}(\mathcal{A})$ be such that $[\mathbf{P},\mathbf{Q}]\in\mathcal{L}$, 
where $\mathcal{L}$ is a subspace of $\mathcal{A}^{s}$. Let $p,q\geq
b_{F}\equiv\min(\max(n_0,0),n+n_0-1)$, where $p\equiv\mathop{\rm
ford}\nolimits\mathbf{P}$,
$q\equiv\mathop{\rm ford}\nolimits\mathbf{Q}$.
Suppose that 
$\mathop{\rm wt}\nolimits(\mathbf{G})\neq(p+q)\alpha/n$ for all
(homogeneous)
$\mathbf{G}\in
S_{F}^{(n_0-1)}(\mathcal{A})\cap\mathrm{Ann}_{F}(\mathcal{A})\cap\mathcal{L}$. 
Then $[\mathbf{P},\mathbf{Q}]=0$.
\looseness=-1
\end{sergy-cor}
\noindent{\em Proof.}  
If $\mathbf{P},\mathbf{Q}\in\mathrm{Ann}_{F}(\mathcal{A})$,
$[\mathbf{P},\mathbf{Q}]\in\mathcal{A}^{s}$, $p,q\geq b_F$,
then, using (\ref{sergy-leadfs2}), (\ref{sergy-leadfs2a}) and
(\ref{sergy-leadcomm1}), we find that
$[\mathbf{P},\mathbf{Q}]\in
\mathcal{N}\equiv
S_{F}^{(p+q-1)}(\mathcal{A})\cap\mathrm{Ann}_{F}(\mathcal{A})$.
Eqs.~(\ref{sergy-leadfs2}) or (\ref{sergy-leadfs2a}) for
$\mathfrak{R}=\mathbf{G}'$ 
imply $\mathop{\rm
wt}\nolimits(\mathbf{G})=k\alpha/n\neq
\mathop{\rm wt}\nolimits([\mathbf{P},\mathbf{Q}])=(p+q)\alpha/n$
for all homogeneous 
$\mathbf{G}\in\mathcal{N}$ with 
$k\equiv\mathop{\rm ford}\nolimits\mathbf{G}\geq n_0$. 
Hence, under our assumptions $\mathop{\rm
wt}\nolimits(\mathbf{G})\neq(p+q)\alpha/n$
for all homogeneous
$\mathbf{G}\in\mathcal{N}\cap\mathcal{L}\equiv\mathcal{M}$, and thus by
Lemma~\ref{sergy-lem1} $[\mathbf{P},\mathbf{Q}]=0$.
$\square$ \looseness=-2

For instance, for the integrable \cite{sergy-ct} 
equation $u_{t}=D^2(u_{1}^{-1/2})+u_{1}^{3/2}\equiv K$  with $n_0=2$
and $\alpha=3/2$ the space
$S_{K}^{(1)}(\mathcal{A}_{\mathrm{loc}})\cap\mathrm{Ann}_{K}
(\mathcal{A}_{\mathrm{loc}})$
is spanned by 1 and $u_1$, and $\mathop{\rm wt}\nolimits(1),\mathop{\rm
wt}\nolimits(u_1)\leq 1 <\alpha(p+q)/n=(p+q)/2$ for $p,q\geq b_{K}=2$. 
Hence, by Corollary~\ref{sergy-corann} all (homogeneous) time-independent local
generalized symmetries of formal order $p>1$ for this equation commute.
\looseness=-1

Likewise, using Corollary~\ref{sergy-corann}, we can easily show that 
for any $\lambda$-ho\-mo\-ge\-ne\-ous integrable evolution equation with 
$\lambda\geq 0$ from \cite{sergy-wang1} all its
$x,t$-in\-de\-pen\-dent homogeneous local generalized symmetries of formal order
$k>0$ commute.\looseness=-2

If $n_0\leq 0$
and, in addition to the conditions of Corollary~\ref{sergy-corann} for
$\mathbf{P}$ and $\mathbf{Q}$, 
the commutator $[\mathbf{P},\mathbf{Q}]\in S_{F}(\mathcal{A}_{\mathrm{loc}})$,
$[\mathbf{P},\mathbf{Q}]$ is
$x,t$-independent and $\mathop{\rm
wt}\nolimits([\mathbf{P},\mathbf{Q}])\neq 0$, then
$[\mathbf{P},\mathbf{Q}]=0$. The weight-related conditions are
automatically satisfied, as the only
$x,t$-independent symmetries in
$S_{F}^{(n_0-1)}(\mathcal{A}_{\mathrm{loc}})$ are constant ones, and 
their weight is zero.
\looseness=-2
In particular, for {\em any} homogeneous (with $\alpha\neq 0$)
NWD system of the form $\mathbf{u}_{t}=\Phi(x)\mathbf{u}_{n}+\Psi(x,t)\mathbf{u}_{n-1}+
\mathbf{f}(x,t,\mathbf{u},\dots,\mathbf{u}_{n-2})$, where $\Phi$, $\Psi$ are
$s\times s$ matrices, {\em all} homogeneous $x,t$-in\-de\-pen\-dent local
generalized symmetries of formal order $k>0$ commute.
\looseness=-3

Let $\mathfrak{R}\in FS_{F}^{(2)}(\mathcal{A})$
be a nondegenerate formal symmetry for (\ref{sergy-eveq})--(\ref{sergy-omt}),
$r\equiv\deg\mathfrak{R}\neq 0$. Then by 
(\ref{sergy-leadfs2a})
$\mathfrak{R}=\Gamma^{-1}h(t)\Lambda^{r/n}\Gamma D^{r}+\cdots$, where
$h(t)\equiv\mathop{\rm diag}\nolimits(h_{1}(t),\dots,h_{s}(t))$ 
is a $s\times s$ diagonal matrix.
Assume that $h(t)$ is homogeneous w.r.t.\ $\mathbf{D}$ and
$\zeta_{\mathfrak{R}}\equiv (\alpha/n+
\mathop{\rm wt}\nolimits(h(t))/r)\neq 0$.
Let $\mathrm{Z}_{F,\mathfrak{R}}(\mathcal{A})=\{\mathbf{G}\in
\mathrm{S}_{F}(\mathcal{A})\mid k\equiv\mathop{\rm ford}\nolimits\mathbf{G}\geq n_0;
\mbox{there exists a diagonal}$ $\mbox{matrix}\,c(t),\,
\mathop{\rm wt}\nolimits(c(t))=0,\,
\mbox{such
that}\,\mathbf{G}'=\Gamma^{-1} c(t) (h(t))^{k/r}\Lambda^{k/n}\Gamma D^{k}+\cdots\}$.
We set here $(h(t))^{k/r}\equiv\mathop{\rm
diag}\nolimits((h_{1}(t))^{k/r},\dots,(h_{s}(t))^{k/r})$. 
Let also 
$\mathrm{St}_{F,\mathfrak{R}}(\mathcal{A})=\{\mathbf{G}\in
\mathrm{Z}_{F,\mathfrak{R}}(\mathcal{A})\mid c(t)\,$ $\mbox{is a 
constant\allowbreak\ matrix}\}$,
and $\mathrm{N}_{F,\mathfrak{R}}^{(j)}(\mathcal{A})$ be the set of symmetries
$\mathbf{G}\in S_{F}(\mathcal{A})$ such that $k\equiv\mathop{\rm
ford}\nolimits\mathbf{G}\geq n_0$, $k\leq j$, and 
$\mathbf{G}'=\Gamma^{-1} c(t) (h(t))^{k/r}\Lambda^{k/n}\Gamma D^{k}+\cdots$, 
where $c(t)$ is an $s\times s$ diagonal matrix, different for different
$\mathbf{G}$ and $k$, and the entries of $c(t)$ are linear combinations
of functions of $t$, say, $\psi_{b}(t)$, such that for all $b$ we have 
$\mathop{\rm wt}\nolimits(\psi_{b}(t))<\zeta_{\mathfrak{R}}(j-k)$  for
$\zeta_{\mathfrak{R}}>0$ and $\mathop{\rm
wt}\nolimits(\psi_{b}(t))>\zeta_{\mathfrak{R}}(j-k)$ for $\zeta_{\mathfrak{R}}<0$. 
For any homogeneous
$\mathbf{G}\in\mathrm{N}_{F,\mathfrak{R}}^{(j)}(\mathcal{A})$ we have
$\mathop{\rm wt}\nolimits(\mathbf{G})<j\zeta_{\mathfrak{R}}$ for
$\zeta_{\mathfrak{R}}>0$ and 
$\mathop{\rm wt}\nolimits(\mathbf{G})>j\zeta_{\mathfrak{R}}$ for
$\zeta_{\mathfrak{R}}<0$, so $\mathop{\rm wt}\nolimits(\mathbf{H})\neq\mathop{\rm
wt}\nolimits(\mathbf{P})$
for any homogeneous $\mathbf{P}\in
\mathrm{Z}_{F,\mathfrak{R}}(\mathcal{A})$ and $\mathbf{H}\in
\mathrm{N}_{F,\mathfrak{R}}^{(\mathop{\rm ford}\nolimits\mathbf{P})}(\mathcal{A})$.
\looseness=-2

Let $\mathbf{P},\mathbf{Q}\in S_{F}(\mathcal{A})$ be homogeneous, and
$[\mathbf{P},\mathbf{Q}]\in\mathcal{L}_{1}\cup\mathcal{L}_{2}$,
where $\mathcal{L}_{1}$ is a subspace of 
$\mathrm{N}_{F,\mathfrak{R}}^{(j)}(\mathcal{A})$
for some $j$ and $\mathfrak{R}$, and ${\mathcal{L}}_{2}$ is a subspace 
of $S_{F}^{(d)}(\mathcal{A})$ for some $d$. Assume that $\mathfrak{R}$
satisfies the above conditions, $\mathop{\rm
wt}\nolimits([\mathbf{P},\mathbf{Q}])\geq j\zeta_{\mathfrak{R}}$ for
$\zeta_{\mathfrak{R}}>0$ and
$\mathop{\rm wt}\nolimits([\mathbf{P},\mathbf{Q}])\leq j\zeta_{\mathfrak{R}}$ 
for $\zeta_{\mathfrak{R}}<0$, and $\mathop{\rm wt}\nolimits(\mathbf{H})\neq
\mathop{\rm wt}\nolimits([\mathbf{P},\mathbf{Q}])$ for all
(homogeneous) $\mathbf{H}\in\mathcal{L}_{2}/(\mathcal{L}_{2}\cap
\mathrm{N}_{F,\mathfrak{R}}^{(j)}(\mathcal{A}))$. Then by Lemma~\ref{sergy-lem1}
$[\mathbf{P},\mathbf{Q}]=0$.\looseness=-1
\looseness=-2

Suppose that, in
addition to the above conditions for $[\mathbf{P},\mathbf{Q}]$, we have 
$d<0$, $\mathop{\rm wt}\nolimits([\mathbf{P},\mathbf{Q}])>0$ for
$\zeta_{\mathfrak{R}}>0$ and $\mathop{\rm wt}\nolimits([\mathbf{P},\mathbf{Q}])<0$
for $\zeta_{\mathfrak{R}}<0$,
and $[\mathbf{P},\mathbf{Q}]$ belongs to
$S_{F}(\mathcal{A}_{\mathrm{loc}})$ and 
can be represented (as function of $t$ and $x$) 
as a polynomial in variables $\chi(t)$ and $\xi(x)$ such that
$\mathop{\rm wt}\nolimits(\chi(t))<0$ and $\mathop{\rm wt}\nolimits(\xi(x))<0$ for
$\zeta_{\mathfrak{R}}>0$, and $\mathop{\rm
wt}\nolimits(\chi(t))>0$ and $\mathop{\rm
wt}\nolimits(\xi(x))>0$ for $\zeta_{\mathfrak{R}}<0$. 
Then $[\mathbf{P},\mathbf{Q}]=0$,  
and there is no further weight-related conditions to verify. 
Indeed, $S_{F}^{(d)}(\mathcal{A}_{\mathrm{loc}})$ for any $d<0$ is
spanned by the symmetries of the form $\mathbf{G}=\mathbf{G}(x,t)$, and for any
homogeneous symmetry $\mathbf{H}=\mathbf{H}(x,t)$ being a polynomial in 
$\chi(t)$ and $\xi(x)$ we obviously have $\mathop{\rm
wt}\nolimits(\mathbf{H})\neq\mathop{\rm
wt}\nolimits([\mathbf{P},\mathbf{Q}])$.

Note that under the assumptions of Proposition~\ref{sergy-prop3} all
$\mathbf{G}\in S_{F}(\mathcal{A})$ with 
$\mathop{\rm ford}\nolimits\mathbf{G}\geq m+n+n_0$ belong to
$\mathrm{St}_{F,\mathfrak{R}}(\mathcal{A})$ by Corollary~\ref{sergy-cor2}.
Suppose that $\mathfrak{R}$
satisfies the conditions, given above. 
Let $d=\min(m+n+n_0-1,p+q)$. 
Then 
for any 
$\mathbf{P},\mathbf{Q}\in S_{F}(\mathcal{A})$ 
such that
$[\mathbf{P},\mathbf{Q}]\in\mathcal{A}^{s}$ we 
have
$[\mathbf{P},\mathbf{Q}]\in
\mathrm{N}_{F,\mathfrak{R}}^{(p+q)}(\mathcal{A})\cup
S_{F}^{(d)}(\mathcal{A})$. 
Then 
$[\mathbf{P},\mathbf{Q}]=0$ for homogeneous
$\mathbf{P},\mathbf{Q}\in \mathrm{Z}_{F,\mathfrak{R}}(\mathcal{A})$, once 
$\mathop{\rm wt}\nolimits(\mathbf{H})\neq\mathop{\rm
wt}\nolimits([\mathbf{P},\mathbf{Q}])$ 
for all (homogeneous) $\mathbf{H}\in
S_{F}^{(d)}(\mathcal{A})/(S_{F}^{(d)}(\mathcal{A})\cap
\mathrm{N}_{F,\mathfrak{R}}^{(p+q)}(\mathcal{A}))$. 
If $p,q>n+n_0-2$, then by (\ref{sergy-leadcomm1})
we can take $d=\min(m+n+n_0-1,p+q-n)$
(or $d=\min(m+n+n_0-1,p+q-n-1)$, if 
$p c_{p}(t)\dot d_{q}(t)-q d_{q}(t)\dot c_{p}(t)=0$).
\looseness=-2

If $\partial\mathbf{F}/\partial t=\partial X_{\gamma}/\partial t=\partial
T_{\gamma}/\partial t=0$, $\gamma=1,\dots,c$, then $\mathbf{F}\in
S_{F}(\mathcal{A})$, and $\partial\mathbf{P}/\partial t=[\mathbf{P},\mathbf{F}]\in  
S_{F}^{(p)}(\mathcal{A})$ for $\mathbf{P}\in S_{F}(\mathcal{A})$. So, taking
$\mathbf{Q}=\mathbf{F}$ and imposing the extra condition $d\leq p$ in three 
previous paragraphs yields valid results. 
\looseness=-1 

We also have the following 
\begin{sergy-prop}\label{sergy-prop2}
Let 
$\alpha\neq 0$ and
$\partial\mathbf{F}/\partial t=0$, $\partial X_{\gamma}/\partial t=\partial
T_{\gamma}/\partial t=0$, $\gamma=1,\dots,c$;  let homogeneous
$\mathbf{P}\in S_{F}(\mathcal{A})$ be such that
$p\equiv\mathop{\rm ford}\nolimits\mathbf{P}\geq n_0$, $\mathop{\rm
ford}\nolimits\partial\mathbf{P}/\partial t<p$ and
$[\mathbf{P},\mathbf{F}]\in\mathcal{L}$, where $\mathcal{L}$ is a
subspace of $\mathcal{A}^{s}$. 
Suppose that 
$\mathop{\rm wt}\nolimits(\mathbf{G})\neq(p+n)\alpha/n$ for all (homogeneous)
$\mathbf{G}\in S_{F}^{(p-1)}(\mathcal{A})\cap\mathcal{L}$
such that
$\mathbf{G}\not\in\mathrm{N}_{F,\mathbf{F}'}^{(p+n)}(\mathcal{A})$. 
Then $[\mathbf{P},\mathbf{F}]=0$, and thus $\partial\mathbf{P}/\partial t=0$ and
$\mathbf{P}\in\mathrm{Ann}_{F}(\mathcal{A})$.\looseness=-1
\end{sergy-prop}
{\em Proof.}
As 
$\mathop{\rm ford}\nolimits\partial\mathbf{P}/\partial t<p$,
we have 
$\partial\mathbf{P}/\partial t=[\mathbf{P},\mathbf{F}]\in 
S_{F}^{(p-1)}(\mathcal{A})\cap\mathcal{L}\equiv \mathcal{M}$. The conditions
$\mathop{\rm ford}\nolimits\partial\mathbf{P}/\partial t<p$ and $p\geq n_0$
by virtue of (\ref{sergy-leadfs2}) or (\ref{sergy-leadfs2a})
for $\mathfrak{R}=\mathbf{P}'$  
readily imply  
$\mathop{\rm wt}\nolimits(\mathbf{P})=p\alpha/n$. Hence
$\mathop{\rm wt}\nolimits([\mathbf{P},\mathbf{F}])=(p+n)\alpha/n$, and thus
by Lemma~\ref{sergy-lem1} 
$[\mathbf{P},\mathbf{F}]=0$. $\square$\looseness=-1

Let $\alpha>0$, $\partial\mathbf{F}/\partial t=0$, $\partial
X_{\gamma}/\partial t=\partial T_{\gamma}/\partial t=0$, $\gamma=1,\dots,c$, and 
homogeneous $\mathbf{P},\mathbf{Q}\in\mathrm{St}_{F,\mathbf{F}'}(\mathcal{A})$,
$p,q\geq n_0$, be polynomials~in~$t$. If we take the
space of symmetries from $S_{F}(\mathcal{A})$ polynomial in time $t$,
for~$\tilde{\mathcal{L}}$, and set
$\mathcal{L}_{1}=\mathrm{N}_{F,\mathbf{F}'}^{(p+q)}(\mathcal{A})
\cap\tilde{\mathcal{L}}$,
$\mathcal{L}_{2}=S_{F}^{(n_0-1)}(\mathcal{A})\cap\tilde{\mathcal{L}}$, $d=n_0-1$,
then
$[\mathbf{P},\mathbf{Q}]\in\mathcal{L}_{1}\cup\mathcal{L}_{2}\equiv\mathcal{M}$,
and thus
the weight-related conditions of
Lemma~\ref{sergy-lem1}, Corollary~\ref{sergy-corann},
Proposition~\ref{sergy-prop2}, etc., are to be checked only for
(homogeneous)
$\mathbf{G}\in \mathcal{L}_{2}$.
Furthermore, if $n_0\leq 0$, then 
$S_{F}^{(n_0-1)}(\mathcal{A}_{\mathrm{loc}})$ contains only
the symmetries $\mathbf{G}=\mathbf{G}(x,t)$, and so 
{\em any} homogeneous local generalized symmetry $\mathbf{K}$ 
of formal order 
$k>0$ being polynomial in~$t$ and $x$ and such that 
$\partial^2\mathbf{K}/\partial\mathbf{u}_{k}\partial t=0$ is in fact
time-independent, 
and any two such symmetries commute. 
This result applies e.g.\
to {\em any} homogeneous
NWD system with $\alpha>0$ having the form
$\mathbf{u}_{t}=\Phi(x)\mathbf{u}_{n}+\Psi(x)\mathbf{u}_{n-1}+
\mathbf{f}(x,\mathbf{u},\dots,\mathbf{u}_{n-2})$, where $\Phi$, $\Psi$ are
$s\times s$ matrices.\looseness=-1 
\subsection{Master symmetries of homogeneous NWD systems}\label{sergy-homogen-ms}
\begin{sergy-cor}\label{sergy-cor3}
Let 
$\alpha\neq 0$, $\partial\Phi/\partial t=0$ and 
$\partial X_{\gamma}/\partial t
=\partial T_{\gamma}/\partial t=0$, $\gamma=1,\dots,c$.
Suppose that there exist a homogeneous $\mathbf{Q}\in\mathop{\rm
Ann}\nolimits_{F}(\mathcal{A})$
and a homogeneous 
$\boldsymbol{\tau}\in\mathcal{A}^{s}$
such that $\partial\boldsymbol{\tau}/\partial
t=0$, $\partial[\boldsymbol{\tau},\mathbf{F}]/\partial t=0$,
$\mathbf{K}=\boldsymbol{\tau}+t[\boldsymbol{\tau},\mathbf{F}]\in
S_F(\mathcal{A})$, $q\equiv\mathop{\rm ford}\nolimits\mathbf{Q}>n+n_0-2$, 
$b\equiv\mathop{\rm
ford}\nolimits[\boldsymbol{\tau},\mathbf{F}]>\max(\mathop{\rm
ford}\nolimits\boldsymbol{\tau},n)$, 
the formal series $([\boldsymbol{\tau},\mathbf{F}])'$ is nondegenerate,
$[[\boldsymbol{\tau},\mathbf{F}],\mathbf{Q}]\in\mathcal{L}$,
where $\mathcal{L}$ is a 
subspace of $\mathcal{A}^{s}$, 
$[\boldsymbol{\tau},\mathbf{Q}]\in\mathcal{A}^{s}$. 
Let $\mathop{\rm wt}\nolimits(\mathbf{H})\neq(b+q)\alpha/n$  
for all (homogeneous)
$\mathbf{H}\in \mathcal{L}\cap
S_{F}^{(n_0-1)}(\mathcal{A})\cap\mathop{\rm
Ann}\nolimits_{F}(\mathcal{A})$. Then
$\mathbf{Q}_{1}=[\boldsymbol{\tau},\mathbf{Q}]\in\mathop{\rm
Ann}\nolimits_{F}(\mathcal{A})$,
and $\mathop{\rm ford}\nolimits\mathbf{Q}_1>q$.\looseness=-1
\end{sergy-cor}
{\em Proof.} From (\ref{sergy-sym}) with $\mathbf{G}=\mathbf{K}$ it clear that 
$[\boldsymbol{\tau},\mathbf{F}]\in\mathrm{Ann}_{F}(\mathcal{A})$, so by 
Corollary~\ref{sergy-corann}
$[[\boldsymbol{\tau},\mathbf{F}],\mathbf{Q}]=0$, whence, using
$[\mathbf{F},\mathbf{Q}]=0$ and the Jacobi identity, we find that $[\mathbf{F},
[\boldsymbol{\tau},\mathbf{Q}]]=0$, so
$[\boldsymbol{\tau},\mathbf{Q}]\in\mathrm{Ann}_{F}(\mathcal{A})$.
By (\ref{sergy-leadcomm1}) the nondegeneracy of 
$([\boldsymbol{\tau},\mathbf{F}])'$ readily
implies 
$\mathop{\rm ford}\nolimits[\boldsymbol{\tau},\mathbf{Q}]=\mathop{\rm
ford}\nolimits[\mathbf{K},\mathbf{Q}]=b+q-n>q$. $\square$
\looseness=-2
\begin{sergy-theo}\label{sergy-th1}
Let the conditions of Corollary \ref{sergy-cor3} be satisfied,  
$\mathop{\rm ad}\nolimits_{[\boldsymbol{\tau},
\mathbf{F}]}^{j}(\mathbf{Q})\in\mathcal{L}_{j}$, where $\mathcal{L}_{j}$
are some subspaces of $\mathcal{A}^{s}$, $\mathbf{Q}_{j}\equiv\mathop{\rm ad}
\nolimits_{\boldsymbol{\tau}}^{j}(\mathbf{Q})\in\mathcal{A}^{s}$, 
and $\mathop{\rm
wt}\nolimits(\mathbf{H})\neq((b-n)j+q+n)\alpha/n$  
for all (homogeneous)
$\mathbf{H}\in \mathcal{L}_{j}\cap
S_{F}^{(n_0-1)}(\mathcal{A})\cap\mathop{\rm
Ann}\nolimits_{F}(\mathcal{A})$, $j=2,\dots,i$.
Then $\mathbf{Q}_{j}
\in\mathop{\rm
Ann}\nolimits_{F}(\mathcal{A})$ and $\mathop{\rm
ford}\nolimits\mathbf{Q}_{j}>\mathop{\rm
ford}\nolimits\mathbf{Q}_{j-1}$, $j=1,\dots,i$.\looseness=-2
\end{sergy-theo}

The proof of the theorem consists in replacing in Corollary \ref{sergy-cor3} the
symmetry
$\mathbf{Q}$ by
$\mathbf{Q}_{j}=\mathop{\rm ad}\nolimits_{\boldsymbol{\tau}}^{j}(\mathbf{Q})$
and repeated use of this corollary for $j=2,\dots,i$.
Note that we can easily verify that 
$\mathop{\rm ad}\nolimits_{\boldsymbol{\tau}}^{j}(\mathbf{Q})\in\mathcal{A}^{s}$,
using Proposition~\ref{sergy-prop4}.
\looseness=-1

Thus, Proposition~\ref{sergy-prop4}, Corollary~\ref{sergy-cor3} and
Theorem~\ref{sergy-th1}  enable us to ensure that $\boldsymbol{\tau}$ indeed is a
nontrivial master symmetry, producing a sequence of symmetries of infinitely growing
formal orders, without assuming {\em a priori} the existence of hereditary recursion
operator
\cite{sergy-oev} or e.g.\ of ``negative" master symmetries $\boldsymbol\tau_{j}$,
$j<0$ \cite{sergy-dor}. So, our results provide a
useful complement to the known general results on master symmetries, cf.\ e.g.\
\cite{sergy-blaszak,sergy-fu,sergy-oev,sergy-dor}.
\looseness=-1

It is important to stress that in general the symmetries $\mathbf{Q}_i$ are not
obliged to commute pairwise. The check of their commutativity and picking out the
commutative subset in the sequence of $\mathbf{Q}_{i}$ can be performed using either
the results of present paper or other methods, see e.g.\ 
\cite{sergy-o,sergy-fu,sergy-oev,sergy-dor}.\looseness=-1

We often can take $[\boldsymbol{\tau}, \mathbf{F}]$ or
$\mathbf{F}$ for ${\bf Q}$, and then in
order to use Theorem~1 it suffices to know only a suitable 
`candidate' $\boldsymbol\tau$ for the master symmetry.
\looseness=-1

For instance, integrable Harry Dym equation $u_t=u^3 u_3\equiv H$, see
e.g.\ \cite{sergy-o,sergy-ibrbook},
satisfies the conditions of Proposition~\ref{sergy-prop4} and of
Theorem~\ref{sergy-th1} for all
$i=2,3,\dots$ with
$\alpha=3$, $b=5$,
$\mathcal{A}=\mathcal{A}(\Omega_{\mathrm{UAC},H})$,
$\boldsymbol{\tau}=u^3 D^3(u \omega_1)\equiv\boldsymbol{\tau}_{0}+u^3 u_3\omega_1$,
$\boldsymbol{\tau}_{0}\in\mathcal{A}_{\mathrm{loc}}$, 
$\mathbf{Q}=[\boldsymbol\tau,u^3
u_{3}]=3u^{5}u_{5}+\cdots\in\mathop{\rm
Ann}\nolimits_{H}(\mathcal{A})$. In particular, the
nonlocal variable $\omega_1$ in $\boldsymbol{\tau}$ is defined by
means of the relations
$\partial\omega_1/\partial t=-u u_2-u_1^2/2$,
$\partial\omega_1/\partial x=u^{-1}$ (informally, $\omega_1=D^{-1}(u^{-1})$). Thus,
by Theorem~\ref{sergy-th1} ${\bf Q}_{j}=\mathop{\rm
ad}\nolimits_{\boldsymbol{\tau}}^{j}(\mathbf{Q})\in\mathop{\rm
Ann}\nolimits_{H}(\mathcal{A})$,
$j=1,2,\dots$, together with
${\bf Q}_{-1}\equiv u^3u_{3}\in\mathop{\rm
Ann}\nolimits_{H}(\mathcal{A}_{\mathrm{loc}})$ and ${\bf Q}_{0}\equiv{\bf
Q}$ form
the infinite hierarchy of 
time-in\-de\-pen\-dent 
symmetries for the Harry Dym equation. 
The commutativity of 
$\mathbf{Q}_{j}$, $j=-1,0,1,\dots$, readily follows from
Corollary~\ref{sergy-corann}. Note that it is possible to show
that $\mathbf{Q}_{j}$,
$j=0,1,\dots$, are in fact {\em local} generalized symmetries of Harry Dym
equation and coincide with the members of hierarchy generated by means of the
recursion operator $\mathfrak{R}=u^3 D^3\circ u\circ D^{-1}\circ u^{-2}$
from the seed symmetry $u^3u_3$ up to the constant
multiples.\looseness=-1
\section*{Acknowledgements}
I am sincerely grateful to the organizers of NATO ARW {\em
Noncommutative Structures in Mathematics and Physics} for inviting
me to participate and to give a talk and for their hospitality.
It is also my great pleasure to thank 
Profs. M. B\l aszak, B.~Fuchssteiner, B.
Kupershmidt, P. Olver and V.V.~Sokolov
for stimulating discussions and comments. Last but not least, 
I acknowledge with deep gratitude Dr. M. Marvan's reading the 
drafts of this paper and making a lot of precise remarks, which 
considerably improved it.\looseness=-2

This research was supported 
by the Ministry of Education, Youth and Sports of Czech
Republic, Grants CEZ:J10/98:\allowbreak{}192400002 and  VS 96003.
\looseness=-2

\end{document}